\documentclass[iop]{emulateapj}
\usepackage{graphicx}
\usepackage{amsmath, mathrsfs, amssymb}
\usepackage{hyperref}

\newcommand\m{\mathrm}
\newcommand\Yc{\tilde{Y}^\m{cyl}_c}

\shorttitle{Stacked Thermal Sunyaev-Zel'dovich Signal of Locally Brightest Galaxies}
\shortauthors{Greco et al.}

\begin{document}

%%%% title/authors
\title{The Stacked Thermal Sunyaev-Zel'dovich Signal of Locally Brightest Galaxies in Planck Full Mission Data: Evidence for Galaxy Feedback?}
\author{Johnny~P.~Greco$^1$, J.~Colin Hill$^{1,2}$, David~N.~Spergel$^1$, and Nicholas~Battaglia$^{1,3}$}
\affil{\small$^1$Department of Astrophysical Sciences, Princeton University, Princeton, NJ 08544}
\affil{\small$^2$Department of Astronomy, Columbia University, New York, NY 10027}
\affil{\small$^3$McWilliams Center for Cosmology, Department of Physics, Carnegie Mellon University, Pittsburgh, PA 15213}
\email{jgreco@astro.princeton.edu}
\email{jch@astro.columbia.edu}
\email{dns@astro.princeton.edu}
\email{nbatta@astro.princeton.edu}
%%%%


\begin{abstract}
We use the Planck full mission temperature maps to examine the stacked thermal Sunyaev-Zel'dovich (tSZ) signal of 188042 ``locally brightest galaxies'' (LBGs) selected from the Sloan Digital Sky Survey Data Release 7.  Our LBG sample closely matches that of \citet[][PCXI]{PlanckIR_XI}, but our analysis differs in several ways.  We work directly in terms of physically observable quantities, requiring minimal assumptions about the gas pressure profile.  We explicitly model the dust emission from each LBG and simultaneously measure both the stacked tSZ and dust signals as a function of stellar mass $M_*$.  There is a small residual bias in stacked tSZ measurements; we measure this bias and subtract it from our results, finding that the effects are non-negligible at the lowest masses in the LBG sample.  Finally, we compare our measurements with two pressure profile models, finding that the profile from \citet{Battaglia2012} provides a better fit to the results than the \citet{Arnaud2010} ``universal pressure profile''. However, within the uncertainties, we find that the data are consistent with a self-similar scaling with mass --- more precise measurements are needed to detect the relatively small deviations from self-similarity predicted by these models. Consistent with PCXI, we measure the stacked tSZ signal from LBGs with stellar masses down to $\log_{10}(M_*/M_{\odot}) \sim 11.1-11.3$. For lower stellar masses, however, we do not see evidence for a stacked tSZ signal.  We note that the stacked dust emission is comparable to, or larger than, the stacked tSZ signal for $\log_{10}(M_*/M_{\odot}) \lesssim 11.3$. Future tSZ analyses with larger samples and lower noise levels should be able to probe deviations from self-similarity and thus provide constraints on models of feedback and the evolution of hot halo gas over cosmic time.
\end{abstract}

% ***********************************************************************
\section{Introduction}\label{sec:intro}
If gravitational dynamics alone determined the properties of gas in galaxy halos, then there should be an approximately self-similar relation between the gas pressure profile and halo mass \citep[e.g.,][]{Kaiser1986, Komatsu2001}. However, non-gravitational processes such as star formation, supernova, and active galactic nucleus (AGN) feedback; bulk turbulent pressure support; non-equilibrated electrons; cosmic rays; magnetic fields; and plasma physics instabilities are expected to lead to deviations from self-similarity \citep[e.g.,][]{Borgani2004, Rudd2009, Parrish2012, Battaglia2012b, McCourt2013, LeBrun2014, Nelson2014} --- particularly in halos well below the cluster mass scale ($\lesssim 10^{14-14.5} \, M_{\odot}$). Galaxy groups and low-mass clusters are ideal laboratories for testing this expectation since hot gas in the intragroup/intracluster medium (ICM) makes it possible to observe their baryons in both stellar and gaseous phases, and their relatively shallow potential wells (compared to very massive clusters) should increase the impact that non-gravitational effects have on their formation and evolution \citep[e.g.,][]{Sijacki2007, Puchwein2008, McCarthy2010}. Furthermore, a significant fraction of the galaxy population resides in small groups \citep[][and references therein]{Mulchaey2000}; hence, observations of these systems can potentially shed light on the physics that dominates galaxy formation and evolution in the universe. 

The thermal Sunyaev-Zel'dovich (tSZ) effect is a measure of the integrated electron pressure along the line-of-sight (LOS) to a galaxy group or cluster. It is, therefore, an excellent tool for studying the thermodynamic state of the ICM. Although high signal-to-noise tSZ observations of small galaxy groups are not feasible with current data, the average gas content of their halos can be studied through statistical stacking methods  \citep[e.g.,][]{Hand2011, PlanckXII, Gralla2014}. Another method for probing the distribution of halo gas, and hence feedback effects, is measuring the cross-correlation between the tSZ and lensing signals \citep[e.g.,][]{VanWaerbeke2014, Hill2014, Battaglia2014, Hojjati2014}. In this work, we use multi-frequency Planck full mission data to stack on the positions of locally brightest galaxies (LBGs). These LBGs are selected from the Sloan Digital Sky Survey Data Release 7 \citep[SDSS/DR7,][]{SDSSDR7}, following the selection criteria of \citet[][hereafter PCXI]{PlanckIR_XI}, which maximizes the fraction of these objects that are the central galaxies of their dark matter halos. This work is intended to be a re-analysis and extension of the study carried out by PCXI. 

The analysis of PCXI suggests a number of potentially unexpected conclusions. Specifically, they detect the tSZ signal from LBGs with stellar masses as low as $M_* = 2\times10^{11}\, M_\odot$, with a clear indication of signal down to $10^{11}\, M_\odot$. They then extract the underlying tSZ signal-halo mass scaling relation from their measurements using mock LBG catalogs derived from the semi-analytic galaxy formation simulation of \citet{Guo2011} --- this simulation uses the technique of \citet{Angulo2010} to rescale the Millennium Simulation \citep{Springel2005} to the WMAP7 cosmology. According to their results, this scaling relation is described by a single power law with no evidence of deviation over the halo mass range of the most massive clusters in the universe ($M_{500}\sim10^{15}\, M_\odot$) down to small groups of galaxies ($M_{500} \sim 4\times10^{12}\, M_\odot$). This remarkable self-similarity in the gas properties of dark matter halos is counterintuitive, as one would naively expect non-gravitational effects such as those mentioned above to play rather different roles in halos over such a wide mass range. Moreover, these results have significant implications for galaxy formation and evolution, as well as the interplay between the baryonic content of galaxies and their parent dark matter halos. It is for these reasons that a re-analysis of the work of PCXI is both interesting and worthwhile as an independent cross-check of their results. 

While our analysis closely follows that of PCXI, it differs in the following significant ways. Whereas PCXI extract the tSZ signal with a multi-frequency matched filter, we use an aperture photometry method (\textsection\ref{sec:extract}), which does not require strong assumptions about the ICM pressure profile and allows us to explicitly treat dust emission from the LBGs and their host halos (note PCXI use aperture photometry to test their primary results). Another significant difference is that we measure and subtract off a stacking induced bias, which arises from the strictly positive and purely additive nature of the Compton-$y$ parameter (\textsection\ref{sec:bias}). This bias has not been accounted for in previous tSZ analyses and is relevant at the lowest mass scales probed by PCXI. Similar to PCXI, we compare our measurements with theoretical predictions based on the ICM pressure profile of \citet{Arnaud2010}, but we also use the pressure profile of \citet{Battaglia2012} to calculate an additional set of predictions (\textsection\ref{sec:tSZE} and \textsection\ref{sec:tSZsig}). Finally, in addition to the default predictions from these pressure profile models, we adjust their parameters to generate purely self-similar predictions, which we use to test the data for deviations from self-similarity~(\textsection\ref{sec:tSZsig}).

Throughout this paper we assume a flat $\Lambda$CDM cosmology consistent with the WMAP9 parameters \citep{Hinshaw2013}. In particular, we assume $\Omega_m=0.272$, $\Omega_\Lambda=0.728$, $\Omega_b = 0.0455$, and $H_0 = 70.4\ \m{km\, s^{-1}\, Mpc^{-1}}$. We approximate the redshift evolution of the Hubble parameter as $H^2(z)/H_0^2 \equiv E^2(z) = \Omega_m (1+z)^3 + \Omega_\Lambda$, which is valid for our redshift range of interest. Cluster parameters are expressed in terms of $\Delta$, where $M_\Delta = \frac{4}{3}\pi R_\Delta^3\, \rho_\m{crit}(z)\Delta$ is the mass enclosed by the radius $R_\Delta$, within which the mean mass density is $\Delta$ times the critical density of the universe, $\rho_\m{crit}(z) = 3 H^2(z) / 8 \pi G$. 

\section{The Thermal SZ Effect}

\subsection{Modeling the ICM Pressure Profile} \label{sec:tSZE}

The tSZ effect is the result of the inverse Compton scattering of cosmic microwave background (CMB) photons off hot electrons. Observationally, it produces a frequency-dependent change in the CMB temperature along the LOS to a galaxy group or cluster. The temperature change at frequency $\nu$ induced by a cluster of mass $M$ at redshift $z$ is given by \citep{SZ1972}
\begin{align}
\frac{\Delta T(\theta, \nu, M, z)}{T_\m{CMB}} &= g(\nu)\, y(\theta, M, z), \nonumber \\
 						    	     &= g(\nu)\, \frac{\sigma_T}{m_e c^2}\int P_e(r, M, z)\,d\ell,
\label{eqn:tSZdef}
\end{align}
where $T_\m{CMB}=2.7255\, K$ is the CMB temperature, $g(\nu) = x \coth(x/2) - 4$ is the tSZ spectral function with $x\equiv h \nu/ k_\m{B} T_\m{CMB}$,  $\theta$ is the angular position with respect to the center of the cluster, $y(\theta, M, z)$ is the standard Compton-$y$ parameter, $\sigma_T$ is the Thomson scattering cross-section, $m_e c^2$ is the rest mass energy of the electron, and $P_e(r, M, z)$ is the electron pressure profile. Here, $r^2=\ell^2 + d_A^2(z) \,\theta^2$, where $d_A(z)$ is the angular diameter distance to the cluster. Equation~(\ref{eqn:tSZdef}) neglects relativistic corrections \citep{Nozawa2006}, which are negligible for essentially all of the systems in our sample, with the possible exception of the highest stellar mass bin. The corrections for this bin would be at most a few percent, and this is well below our statistical errors. Note that throughout this  work we assume the pressure profile is spherically symmetric, which translates into a Compton-$y$ profile that is azimuthally symmetric in the plane of the sky.  

To make theoretical calculations of the tSZ signal, we adopt two models for the electron pressure profile, both of which are based on the generalized NFW profile first proposed by  \citet{Nagai2007}. The first profile is the ``universal pressure profile'' (UPP) of \citet{Arnaud2010}. This profile is derived from a combination of X-ray observations of massive, $z < 0.3$ clusters and the hydrodynamical simulations of \citet{Nagai2007}, \citet{Borgani2004}, and \citet{Piffaretti2008}. These simulations include radiative cooling, star formation, and energy feedback from supernova explosions (they do not include AGN feedback). The X-ray observations define the radial profile for $r<R_{500}$, and the simulations are used to extend the profile beyond $R_{500}$. For our default UPP calculations, we adopt their empirically derived model parameters: $[P_0, c_{500},\allowbreak \gamma, \alpha, \beta, \alpha_p] = [8.061, 1.177, 0.3081, 1.0510, 5.4905, 0.12]$, neglecting the weak radial dependence of $\alpha_p$, as it introduces an insignificant correction \citep{Arnaud2010}. In addition, we use their standard self-similar model, which is presented in their Appendix~B, to test the consistency of our results with a purely self-similar pressure profile. The normalization of the UPP is obtained using X-ray mass measurements that assume hydrostatic equilibrium (HSE), which simulations suggest are biased low by $\sim\!10-30\%$ \citep[e.g.,][]{Evrad1990, Rasia2006, Nagai2007b, Lau2009, Battaglia2012b, Nelson2012, Rasia2012, Nelson2014} and observations confirm \citep[e.g.,][]{PlanckXX, Hill2014, von_der_Linden2014, Simet2015}. Therefore, we assume a ``hydrostatic mass bias'' of $(1-b)=0.8$, where $M^\m{HSE}_{500}=(1-b)M_{500}$, for both our default and self-similar UPP calculations.

The second profile we adopt is the parameterized fitting function of the ICM pressure profile given in \citet[][hereafter the Battaglia et al. pressure profile or BPP]{Battaglia2012}, which is derived from the cosmological hydrodynamics simulations of \citet{Battaglia2010}. These simulations were run using the smooth particle hydrodynamic code GADGET-2 \citep{Springel2005b} and include radiative cooling and sub-grid prescriptions for star formation, supernova feedback, and AGN feedback. Further, the smoothed particle hydrodynamics used in these simulations naturally captures the effects of non-thermal pressure support due to bulk motions and turbulence, which must be modeled in order to accurately characterize the cluster pressure profile at large radii. Although the BPP is derived solely from numerical simulations, we note that it is in good agreement with a number of observations of cluster pressure profiles, including those based on the REXCESS X-ray sample of massive, $z < 0.3$ clusters \citep{Arnaud2010}, independent studies of low-mass groups at $z < 0.12$ with Chandra \citep{Sun2011}, early Planck measurements of the stacked pressure profile of $z < 0.5$ clusters \citep{Planck_V}, and recent X-ray measurements of high-$z$ cluster pressure profiles \citep{McDonald2014}. Our default BPP calculations use the parameters of their ``AGN feedback,'' $\Delta=200$ model. We also calculate purely self-similar predictions with this model by setting $\alpha_m = \alpha_z=0$ for all the model parameters, where non-zero $\alpha_m$ and $\alpha_z$ describe the mass- and redshift-dependent deviations from self-similarity in the BPP (see Eqn.~(11) in \citet{Battaglia2012}). 

\vspace{0.1cm}
\subsection{The Integrated tSZ Signal} \label{sec:integratedtSZ}

We quantify the tSZ signal as the Compton-$y$ parameter integrated over the solid angle of the cluster, $ d \Omega = dA / d_A^2(z)$, expressed in arcmin$^2$:
\begin{equation}\label{eqn:integratedY}
Y = \int y\, d \Omega = d_A^{-2}(z)\,\frac{\sigma_T}{m_e c^2} \int dA \int d \ell\,\, P_e(r, M, z).
\end{equation}
This quantity is a measure of the total thermal energy of the cluster and is thus expected to correlate strongly with cluster mass \citep[e.g.,][]{Motl2005, Nagai2006, Stanek2010}.  In practice, observations are sensitive to the tSZ signal within some aperture and integrated along the LOS, corresponding to a cylindrical volume of integration in Equation~(\ref{eqn:integratedY}). Hence, the observationally relevant integrated Compton-$y$ parameter is given by 
\begin{equation}\label{eqn:Yc}
Y^\m{cyl}_c = d_A^{-2}(z)\,\frac{\sigma_T}{m_e c^2} \int_0^{R_c}\!\!2\pi R\, d R \int_{-\infty}^\infty \!\!d \ell  \, P_e(r, M, z), 
\end{equation}
where $R$ is the projected radius, and we have truncated the radius of the cylinder at $R_c \equiv 5\times R_{500}$, which lies beyond the virial radius. The exact choice of $R_c$ is arbitrary; the value we adopt was chosen for consistency with the literature \citep[e.g.,][]{Melin2011, PlanckE_X}. 

For comparison with previous work, we also define the Compton-$y$ parameter integrated over a sphere of radius $R_{500}$:
\begin{equation}\label{eqn:Y500}
Y^\m{sph}_{500} = d_A^{-2}(z)\,\frac{\sigma_T}{m_e c^2} \int_0^{R_{500}}\!\! 4\pi\, P_e(r, M, z)\, r^2 dr. 
\end{equation}
We numerically solve for the conversion factor between the cylindrically and spherically integrated Comptonization parameters by assuming the UPP is valid over the entire mass range relevant to this study. Using the model parameters specified in \textsection\ref{sec:tSZE}, we convert $Y^\m{cyl}_c$ into $Y^\m{sph}_{500}$ via the ratio $Y^\m{sph}_{500}/Y^\m{cyl}_{c}=0.5489$. Note this ratio is independent of mass and redshift; however, the BPP predicts that it is a function of both these quantities. 

% Nick's Ycyl to Ysph scatter figure
\begin{figure}[t!]
\begin{center}
\includegraphics[width=9cm, trim=0.3cm 0.cm -1cm -0.1cm,clip=true]{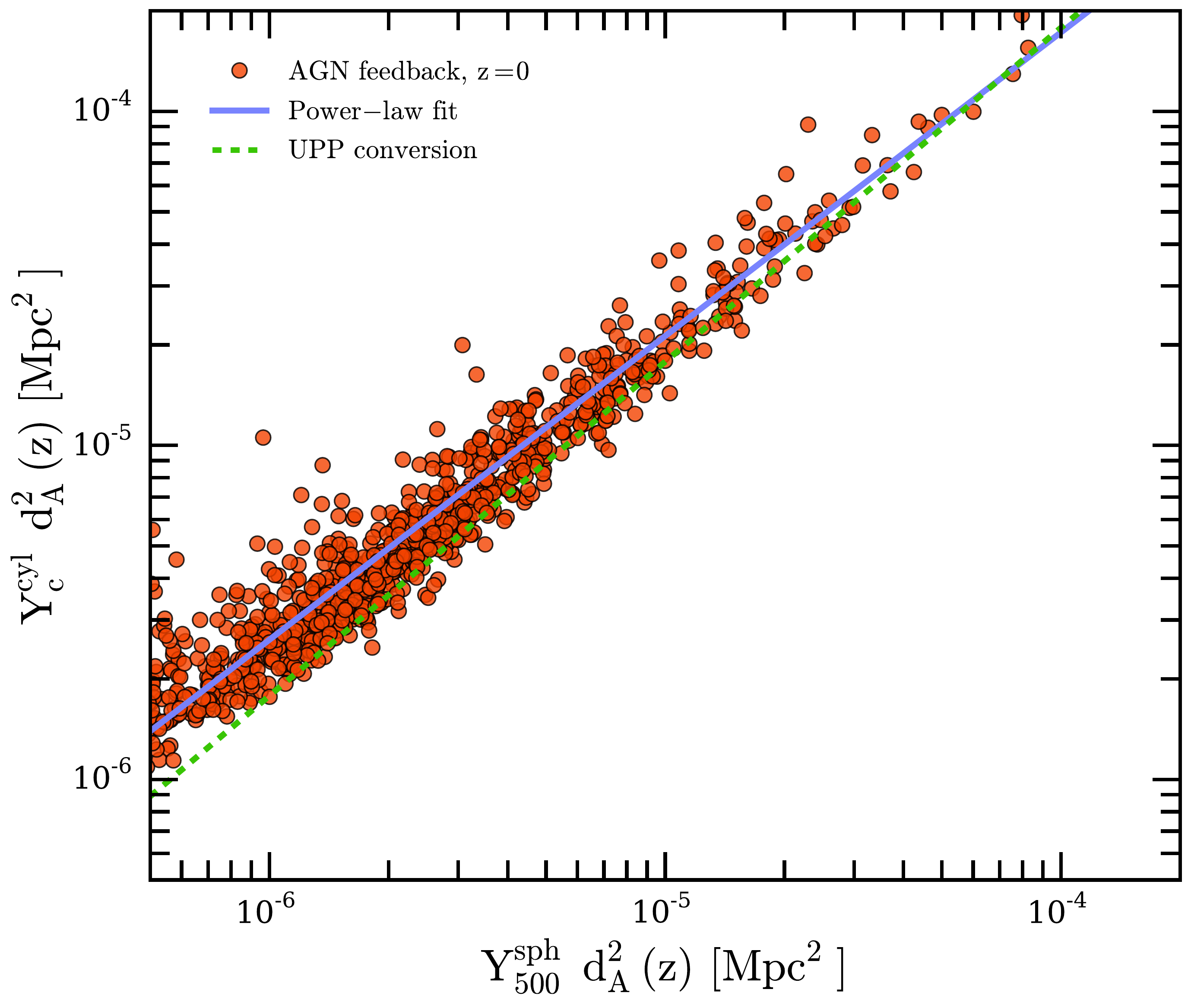}
\caption{The relationship between  $Y^\m{cyl}_c$ and $Y^\m{sph}_{500}$ (Eqn. (\ref{eqn:Yc}) and (\ref{eqn:Y500})) for all halos with $M_{200} \geq 5\times10^{13}\,M_\odot$ at $z=0$ in the \citet{Battaglia2010} ``AGN feedback'' simulations. Red points indicate individual measurements, the dashed green line shows the (mass and redshift independent) UPP prediction, and the solid blue line is a power-law fit. For the power-law fit, each halo is weighted by $Y^\m{sph}_{500}$ so that the low-mass halos do not drive the parameter values. Note the UPP predicts a different slope for the correlation. The intrinsic scatter in the conversion between these quantities --- which has been neglected in previous tSZ analyses --- is 24\% for halos with $M_{200} \geq 5\times10^{13}\,M_\odot$.}
\label{fig:scatter} 
\end{center}
\end{figure}
%%%%
 
It is important to note that the spherically integrated Compton-$y$ is not an observable quantity, as the $y$ signal in an observed CMB map has already been integrated along the entire LOS --- there is no way to ``cut'' the integrals off at a spherical boundary.  The only physically observable integrated Compton-$y$ signal corresponds to a cylindrical volume of integration, as in Equation~(\ref{eqn:Yc}).  For a sufficiently large choice of boundary, the cylindrical and spherical quantities should converge, as the pressure profile in the far outskirts of a cluster becomes small. However, this can only be checked using simulations or theoretical calculations, since the spherically integrated Compton-$y$ is not directly observable.  To go from the cylindrically integrated to the spherically integrated $y$ signal requires either the assumption of a pressure profile shape or a noisy deprojection from 2D to 3D, which is not feasible at the resolution of current large-scale CMB surveys. Furthermore, this deprojection assumes that no other foreground or background objects intersect the LOS.  Many previous analyses \citep[e.g.,][]{Melin2011, PlanckE_X, Sehgal2013} assume the UPP in order to convert from the cylindrical to spherical tSZ quantities.  However, if the goal of tSZ analyses is to understand the behavior of the gas presure profile, it seems undesirable to make strong \emph{a priori} assumptions about it.  Moreover, directly converting $Y^\m{cyl}_c$ to a spherically integrated quantity neglects the significant scatter seen between these quantities in simulations (see Figure~\ref{fig:scatter}).  For these reasons, we choose to work directly in terms of the physically observable $Y^\m{cyl}_c$, only translating to a spherically integrated quantity (using the UPP) when needed for comparison to other results in the literature. 

Figure~\ref{fig:scatter} shows the relationship between  $Y^\m{cyl}_c$ and $Y^\m{sph}_{500}$ as predicted by the ``AGN feedback'' simulations of  \citet{Battaglia2010}. The dashed green line shows the UPP prediction for the conversion between the two quantities, and the solid blue line is a power-law fit. When we fit the power-law, each halo is weighted by $Y^\m{sph}_{500}$ so that the low-mass halos do not drive the parameter values. Although these quantities are strongly correlated, the UPP predicts a different slope for the correlation, and there is significant scatter that has been neglected when performing this conversion in previous tSZ analyses. Averaging over all halos with $M_{200} \geq 5\times10^{13}\,M_\odot$ at $z=0$, we find a 24\% intrinsic scatter in this conversion. \citet{LeBurn2015} --- who used the simulations described in \citet{LeBrun2014} --- have also shown that assuming the UPP is valid for all masses, redshifts, and cluster-centric radii leads to inaccurate conversions between $Y^\m{cyl}_c$ and $Y^\m{sph}_{500}$, particularly for lower-mass objects. 

Under the assumptions of hydrostatic and virial equilibrium, the simplest self-similar model predicts that the cluster gas temperature and mass satisfy the relation \citep{Kaiser1986}
\begin{equation}\label{eqn:TM}
T \propto M^{2/3} E^{2/3}(z). % add note about M virial mass
\end{equation}
For an isothermal ICM, the integrated tSZ signal and the cluster gas temperature are related by
\begin{equation}\label{eqn:YM}
Y \propto  M\, T\, d_A^{-2}(z).
\end{equation}
Combining Equations (\ref{eqn:TM}) and (\ref{eqn:YM}), we find a reference self-similar scaling relationship:
\begin{equation}
Y\propto M^{5/3}\, E^{2/3}(z)\, d_A^{-2}(z).
\end{equation}
Motivated by the above reasoning, as well as for consistency with previous work, we present our results~as
\begin{equation}\label{eqn:Ytw}
\Yc = Y^\m{cyl}_c\, E^{-2/3}(z)\, \left(\frac{d_A(z)}{500\, \m{Mpc}}\right)^2, 
\end{equation}
with an analogous definition for the spherically integrated $Y^\m{sph}_{500}$.

\section{Data and selection}\label{sec:data}

\subsection{Planck intensity maps}\label{sec:planckdata}

Our analysis is based on the (Zodiacal-corrected) full mission maps from the Planck space mission \citep{PlanckR_I}. We make use of data from the Planck High-Frequency Instrument (HFI) at 100, 143, 217, and 353 GHz. In the approximation of circular Gaussian beams, the effective FWHM values for these channels are 9.66, 7.27, 5.01, and 4.86 arcmin \citep[Table 1 of][]{PlanckR_XXI}. The Planck maps are given in HEALPix\footnote{\url{http://healpix.jpl.nasa.gov}} format, with the HFI channels at a resolution of $N_\m{side}=2048$, corresponding to a typical pixel width of 1.7 arcmin. To avoid severe contamination from Galactic dust and point source emission, we use the procedure of \citet{Hill2014} to apply a mask that removes $\sim50\%$ of the most contaminated sky. 

% **** galaxy distribution figure 
\begin{figure*}[t!]
\begin{minipage}[t!]{0.5\linewidth} 
\centering
\includegraphics[width=8.5cm,trim=0.45cm 0.5cm 1.31cm 0cm,clip=true]{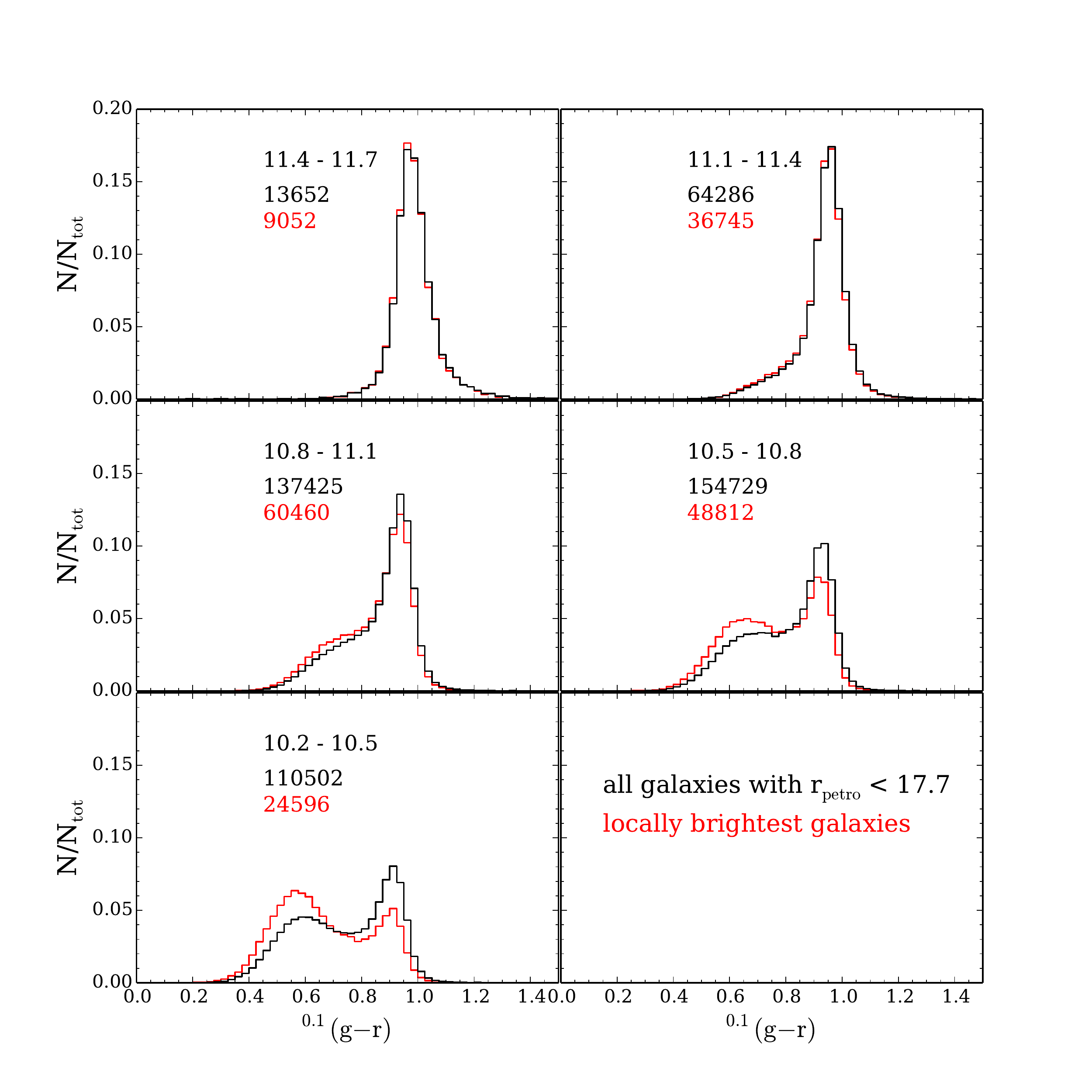}
\end{minipage}
\begin{minipage}[t!]{0.4\linewidth}
\centering
\includegraphics[width=8.5cm,trim=1.31cm 0.5cm 0.5cm 0cm,clip=true]{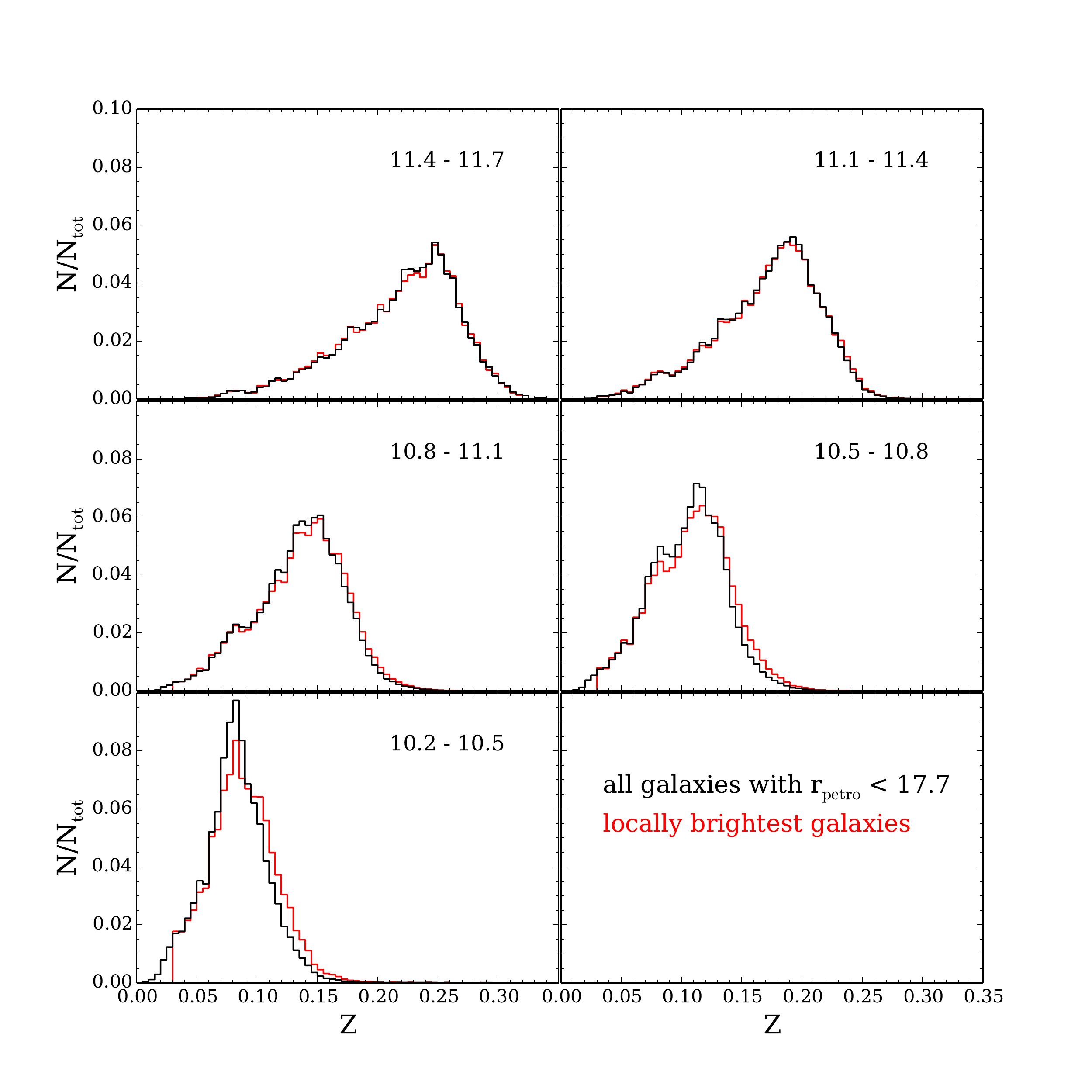}
\end{minipage}
\caption{Comparison of the color (left) and redshift (right) distributions of the parent and LBG samples for 5 different stellar mass bins. The binning and layout of the figure were chosen for easy comparison with Figure 1 of PCXI. The stellar mass ranges are given in $\log_{10}M_*/M_\odot$ at the top of each panel. The additional numbers on the left set of panels indicate the number of galaxies in each bin for the parent and LBG samples. \label{fig:sample}}
\end{figure*}
% **** 

\subsection{NYU-VAGC locally brightest galaxies}\label{sec:LBGs}

We use the New York University Value-Added Catalog\footnote{\url{http://sdss.physics.nyu.edu/vagc}} \citep[NYU-VAGC,][]{Blanton2005} to build a sample of bright central galaxies. The NYU-VAGC is based on a collection of publicly available galaxy catalogs that are cross-matched to the SDSS/DR7. For each galaxy in our sample, this catalog provides positions, magnitudes, spectroscopic redshifts, $K-$corrections, and stellar mass estimates. Details of the derivation of the last two quantities are given in \citet{Blanton2007}.  Briefly, the stellar mass estimates, which are particularly important for our study, are derived from fitting the five-band SDSS photometric data to a large set of spectral templates, which are based on the stellar evolution synthesis models of \citet{Bruzual2003}. These models assume the stellar initial mass function of \citet{Chabrier2003}.  The resulting stellar masses are estimated to have a statistical error of $\sim0.1$~dex \citep{Blanton2007, Li2009}. 

We follow the galaxy selection algorithm of PCXI, which, for completeness, we will summarize here. We start by selecting a sample of galaxies with $r_\m{petro}<17.7$, where $r_\m{petro}$ is the $r-$band, extinction-corrected Petrosian apparent magnitude. This produces a parent sample of 631267 galaxies. Then, following PCXI, we define ``locally brightest galaxies'' (LBGs) to be the subset of galaxies with $z>0.03$ and $r_\m{petro}$ brighter than all other sample galaxies projected within 1.0~Mpc and with $|c\cdot\Delta z|<1000\ \m{km\, s^{-1}}$. This yields a sample of 352216 LBGs. In addition, we restrict the range of stellar masses to $10.0 < \log_{10}(M_*/M_\odot) < 12.0$, reducing the count to 328367. This LBG selection procedure was tested extensively in PCXI, and it appears to be quite robust. In their Appendix~A, they vary the isolation criteria of 1.0~Mpc and $|c\cdot\Delta z|<1000\ \m{km\, s^{-1}}$, finding that such changes do not significantly impact their results. 

For consistency with PCXI, we use the ``photometric redshift 2'' (photoz2) catalog \citep{Cunha2009} to search for galaxies without spectroscopic redshifts whose brightness and proximity with respect to a galaxy in our LBG sample may violate our selection criteria. The photoz2 catalog provides redshift probability distributions for all SDSS galaxies down to much fainter magnitudes than the limits of our sample, and it is available as a value-added catalog on the SDSS/DR7 website\footnote{\url{http://www.sdss.org/dr7/products/value_added}}. Adopting the selection cuts of PCXI, we remove any LBG candidate with a neighbor in this catalog of equal or brighter $r_\m{petro}$, projected within 1.0~Mpc, and with a probability greater than 10\% to have a redshift equal to or less than the candidate. After this filtering process, we are left with 244610 LBGs. Finally, our mask eliminates 56568 galaxies located in regions of significant Galactic dust contamination or very near bright point sources, leaving us with a final sample of 188042 LBGs. Figure~\ref{fig:sample} compares the color and redshift distributions of the parent and LBG samples for 5 different stellar mass bins.  Our LBG sample reproduces the qualitative properties seen in Figure 1 of PCXI. Namely, the distributions are very similar for $\log_{10}(M_*/M_\odot)\geq10.8$, and at lower masses, the LBGs tend to be bluer and at slightly larger redshifts. 

\section{Analysis}

\subsection{Cluster parameters}

We assume each LBG corresponds to the center of a dark matter halo (see \textsection\ref{sec:miscenter} for an empirical assessment of our sensitivity to miscentering). For a given LBG, the NYU-VAGC provides an estimate of its total stellar mass $M_*$, and we use the median $M_*-M_{200}$ relation\footnote{We refer the reader to \textsection2.2.1 of PCXI for a detailed discussion about this stellar-to-halo mass relation.} from Figure 3 of PCXI to convert this into an estimate of the corresponding halo mass $M_{200}$. We then calculate $R_{200}$ with the relation $M_\Delta = \frac{4}{3}\pi R_\Delta^3\, \rho_\m{crit}(z)\Delta$. This radius sets the size of the aperture we use to extract the tSZ signal from each LBG. Note that, since $R \propto M^{1/3}$, our measurements in the $\Yc-M_*$ plane are relatively insensitive to the assumed stellar-to-halo mass relation. 

For easier comparison with PCXI and X-ray results, we convert the cluster parameters to their $\Delta=500$ values. To accomplish this, we assume an NFW \citep{NFW1997} density profile and the concentration parameter, $c_\Delta$, of \citet{Neto2007}. Specifically, we derive a relationship between the $\Delta=200$ and $\Delta=500$ cluster parameters via the following equation:
\begin{equation}
\int_0^{R_\Delta} 4 \pi r^2\, \rho_\m{NFW}(r, M_\Delta, c_\Delta)\, dr = \frac{4}{3}\pi R_\Delta^3\, \rho_\m{crit}(z)\Delta.
\end{equation}

\subsection{Extracting the tSZ signal}\label{sec:extract}

We begin our analysis by smoothing all of the Planck maps to a common resolution of 9.66 arcmin. This value is set by the angular resolution of the 100 GHz map, assuming a Gaussian circular beam. To extract the tSZ signal, we take advantage of its known frequency dependence and the multi-frequency Planck data, which span the null of the tSZ spectral function at $\sim218$ GHz. For each LBG in our sample, we perform the following analysis. 

Consider an LBG whose corresponding halo has an angular radius of $\theta_{500} = R_{500}/d_A(z)$. Centered on this LBG's position, we extract from each HFI map  a circular aperture of angular radius $\theta_c\equiv 5\times\theta_{500}$ and an annular aperture of inner radius $\theta_c$ and outer radius $\theta_c + \m{FWHM}$, where FWHM is the 100~GHz value of 9.66 arcmin (we find that varying the inner radius of the annulus does not significantly impact our results). Next, we define the observed signal in the $i^{th}$ HFI map, $S_i$, to be the sum of all pixels inside the circular aperture minus the mean pixel value inside the annular aperture. This subtraction is meant to remove large-scale foreground contamination, assuming it is roughly constant over the extracted aperture. We then model the observed signal as 
\begin{align}\label{eqn:model}
S_i &= a_i\, Y^\m{cyl}_c + b_i\, D_c + \delta T_\m{CMB},\ \m{where} \\
 a_i &= g(\nu_i)\,T_\m{CMB}, \nonumber \\
b_i &= \left(\frac{\nu_i(1+z)}{\nu_0}\right)^\beta B(\nu_i(1+z), T_\m{dust}) \left[\frac{\partial B(\nu_i, T)}{\partial T}\right]^{-1}_{T_\m{CMB}},\nonumber
\end{align}
$Y^\m{cyl}_c$ is the Comptonization parameter integrated along the full LOS inside a cylinder of radius $R_c=\theta_c\,d_A(z)$, $D_c$ is the amplitude of the integrated dust emission within the same cylinder, $\delta T_\m{CMB}$ accounts for primordial fluctuations in the CMB temperature (which may not average to precisely zero within a finite patch), $\nu_0 = 353$~GHz is the reference frequency, $\beta=1.78$ is the dust spectral emissivity index \citep{Addison2013}, $T_\m{dust}=20$~K is the dust temperature \citep{Draine2011}, and $B(\nu, T)$ is the Planck function. In analogy to Equation~(\ref{eqn:Ytw}), we define the parameter $\tilde{D}_c\equiv D_c\times(d_A(z)/500\ \m{Mpc})^2$, which accounts for the effects of comparing similar objects at different redshifts. We confirm our final results are not significantly influenced by the exact values of $\beta$ and $T_\m{dust}$. We use the bandpass-integrated values of $a_i$ that are given in Table~1 of \citet{PlanckR_XXI}, and we compute $b_i$ by integrating over the $i^{th}$ bandpass \citep{Planck_IX}. Also, the relevant  parameters in Equation (\ref{eqn:model}) are understood to be beam-convolved. 

We use standard least squares fitting to solve for the best fit parameters in Equation~(\ref{eqn:model}). That is, if {\bf M} is the matrix whose $i^{th}$ row corresponds to the $i^{th}$ bandpass and is given by $(a_i, b_i, 1)$, then the best fit parameters are given by
\begin{equation}
\left[({\bf M}^T{\bf M})^{-1} {\bf M}^T\right] \cdot \vec{S}= (Y^\m{cyl}_c,\, D_c,\,\delta T_\m{CMB})^T. 
\end{equation}
To correct  $Y^\m{cyl}_c$ for the effect of the beam, we use the UPP prediction for the ratio $Y^\m{cyl}_c/\allowbreak Y^\m{cyl}_{c,\,b}$, where $Y^\m{cyl}_{c,\,b}$ is the integrated Compton-$y$ parameter convolved with a circular Gaussian beam with FWHM = 9.66 arcmin. This ratio is mass and redshift dependent. Example values of this ratio are $Y^\m{cyl}_c/ Y^\m{cyl}_{c,\,b}$ = 1.61, 1.58, 1.12,  and 1.01 for $\log_{10}(M_*/M_\odot)$ = 10.25, 10.75, 11.25, and 11.75, respectively, where we have assumed the median redshift in our sample for each stellar mass. To test the sensitivity of this ratio to the assumed pressure profile, we compare it with the ratio predicted by a Gaussian profile, with  FWHM equal to that of the corresponding UPP prediction. We find that the predicted ratios differ by less than $5\%$ for the mass and redshift ranges of interest. 

Finally, in \textsection\ref{sec:dust_test}, we test the effectiveness of our dust model by repeating the above analysis with $b_i = 0$. We perform this test two ways: with all four HFI channels used in our fiducial analysis (Analysis~II) and with all channels except the 353~GHz channel (Analysis~III), which is expected to contain the most contamination from dust.

\subsection{Stacking bias}\label{sec:bias}

The Compton-$y$ parameter is strictly positive and purely additive along the LOS. Thus, the stacked tSZ signal from a large number of objects has the potential to be biased high due to background and/or foreground objects, which are not necessarily physically correlated with the objects of interest. Large-scale structure that is correlated with the target objects can also contribute to this bias signal. Additionally, there is a contribution from the non-zero global Compton-$y$ signal arising from reionization, the intergalactic medium, and unresolved massive structures (i.e., galaxies, groups, and clusters). For the most massive clusters, this bias is likely negligible, as these are exceedingly rare objects, and it is very unlikely the LOS will intersect an object that produces a comparable signal. However, when stacking on the positions of low-mass halos, as in the present study, such chance alignments are much more likely, and this may lead to an overestimate of the tSZ signal from these low-mass objects.  If this bias signal is present, it should be evident by stacking on progressively larger numbers of random points, which should produce a signal that asymptotes to a non-zero value.  Similar statements hold true for the dust emission amplitude, $\tilde{D}_c$.  However, the CMB temperature fluctuation, $\delta T_\m{CMB}$, should average to zero. 

% **** mean values
\begin{figure}[t!]
\centering
\includegraphics[width=8.8cm, trim=0.45cm 3cm 1cm 1.8cm,clip=true]{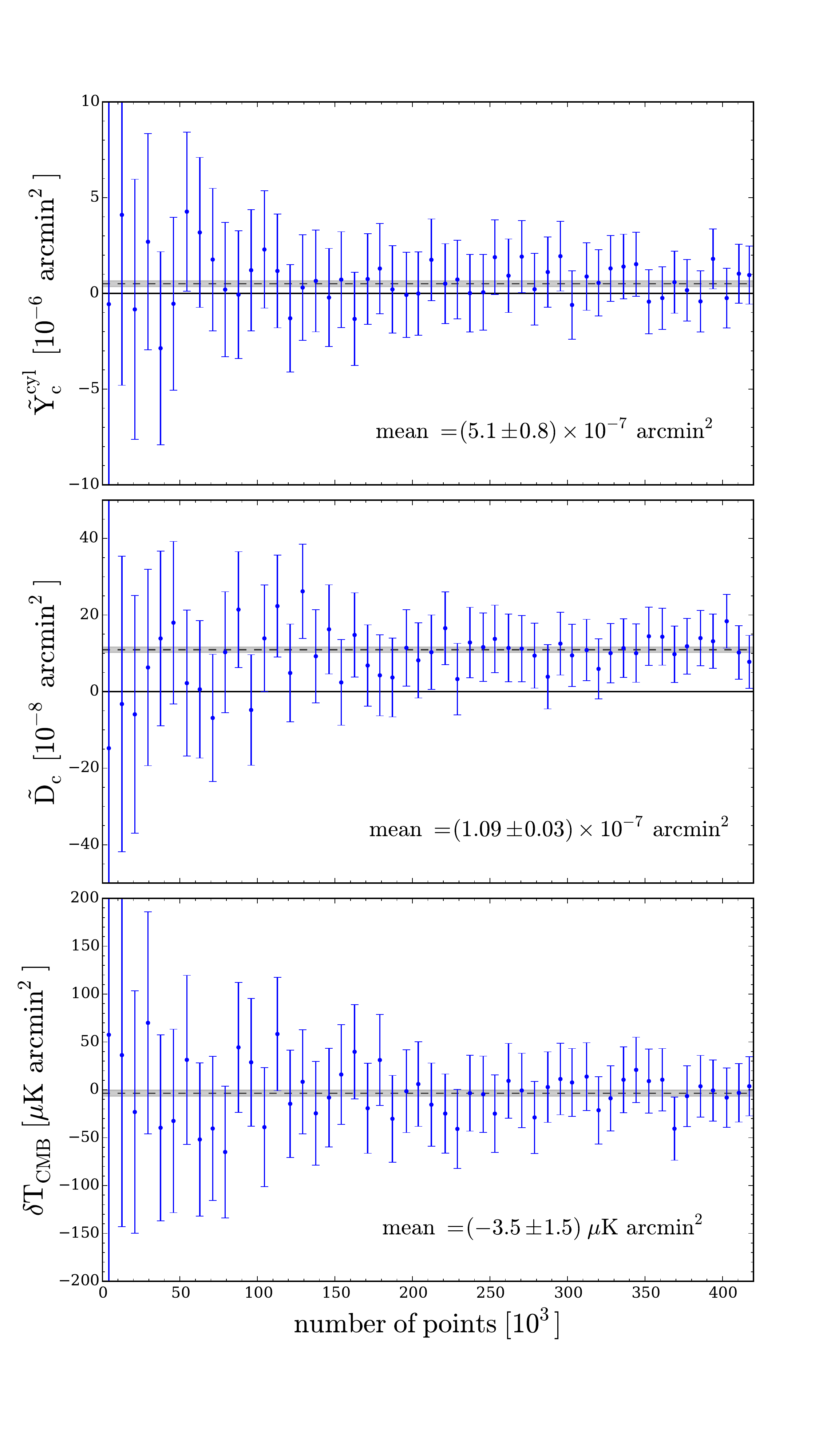}
\caption{Stacking bias test results. Each plot shows mean values of a parameter from our fiducial analysis, where we have stacked on random points in the sky. Error bars show $2\sigma$ uncertainties,  and dashed-lines indicate the inverse-variance weighted averages, with $2\sigma$ errors shown as gray bands. The positive saturation values of $\Yc$ and $\tilde{D}_c$ suggest our measurements are indeed biased high. We subtract the above mean values from all of our analyses. Note the mean values indicated in each panel are quoted with $1\sigma$ uncertainties, which neglect the correlation between points and calibration uncertainties. Note the observed signal is given by Equation~(\ref{eqn:model}), where for the 143~GHz channel, $|a_{143}| = 2.785$~K and $b_{143} \approx 8-12$~K, depending on the redshift to the LBG. \label{fig:mean_vals}}
\end{figure}
%****

With the above statements as our motivation, we search for evidence of this bias signal by running our stacking analysis on random points in the sky, drawing (with replacement) the mass and redshift for each point from their respective LBG distributions. Figure \ref{fig:mean_vals} shows the results, as a function of the number of random points, for each of the parameters in our fiducial analysis. In each panel, error bars represent $2\sigma$ uncertainties, and dashed-lines indicate the inverse-variance weighted averages, with $2\sigma$ errors shown as gray bands. Intriguingly, both $\Yc$ and $\tilde{D}_c$ saturate at positive values, and $\delta T_\m{CMB}$ saturates at a value consistent with zero as expected. Although these saturation values can naively be interpreted as mean signals of the universe, this is not necessarily the case, as their exact values are likely very sensitive to systematic issues such as zero-levels in the maps and calibration uncertainties. It is important to note that one has to consider large numbers of random points to convincingly see this effect, which suggests null tests based on $\lesssim\!10^5$ points may give misleading results. This bias has not been seen (and hence has not been subtracted) in previous tSZ analyses, and our results suggest this may be due stacking on an insufficient number of random points. We subtract the mean values indicated in Figure~\ref{fig:mean_vals} from all our analyses.

We note that the multi-frequency matched filter method employed by PCXI, which effectively
includes a free zero-point level, is not subject to this stacking bias effect (as shown explicitly in \citet{LeBurn2015}). However, as we have shown in this section, aperture photometry methods are subject to this bias. The primary advantage of our aperture photometry method is that it does not require any {\it a priori} assumptions about the ICM pressure profile. Given that the shape and mass/redshift dependence of the pressure profile are not known for low-mass objects, this seems potentially important.

\section{Results}
After running the above analysis on our entire LBG sample, we bin the best-fit parameters by stellar mass and use bootstrap resampling to estimate the binned averages and uncertainties. In this section, we present our stacking results for $\Yc$ and $\tilde{D}_c$, as well as an empirical assessment of our sensitivity to miscentering between the LBGs and their host halos. We note the stacked $\delta T_\m{CMB}$ values are consistent with an average value of zero. 

\subsection{The stacked tSZ signal}\label{sec:tSZsig}

\begin{figure}[t!]
\centering
\includegraphics[width=9.5cm, trim=0.cm 0.cm -0.5cm 0.cm,clip=true]{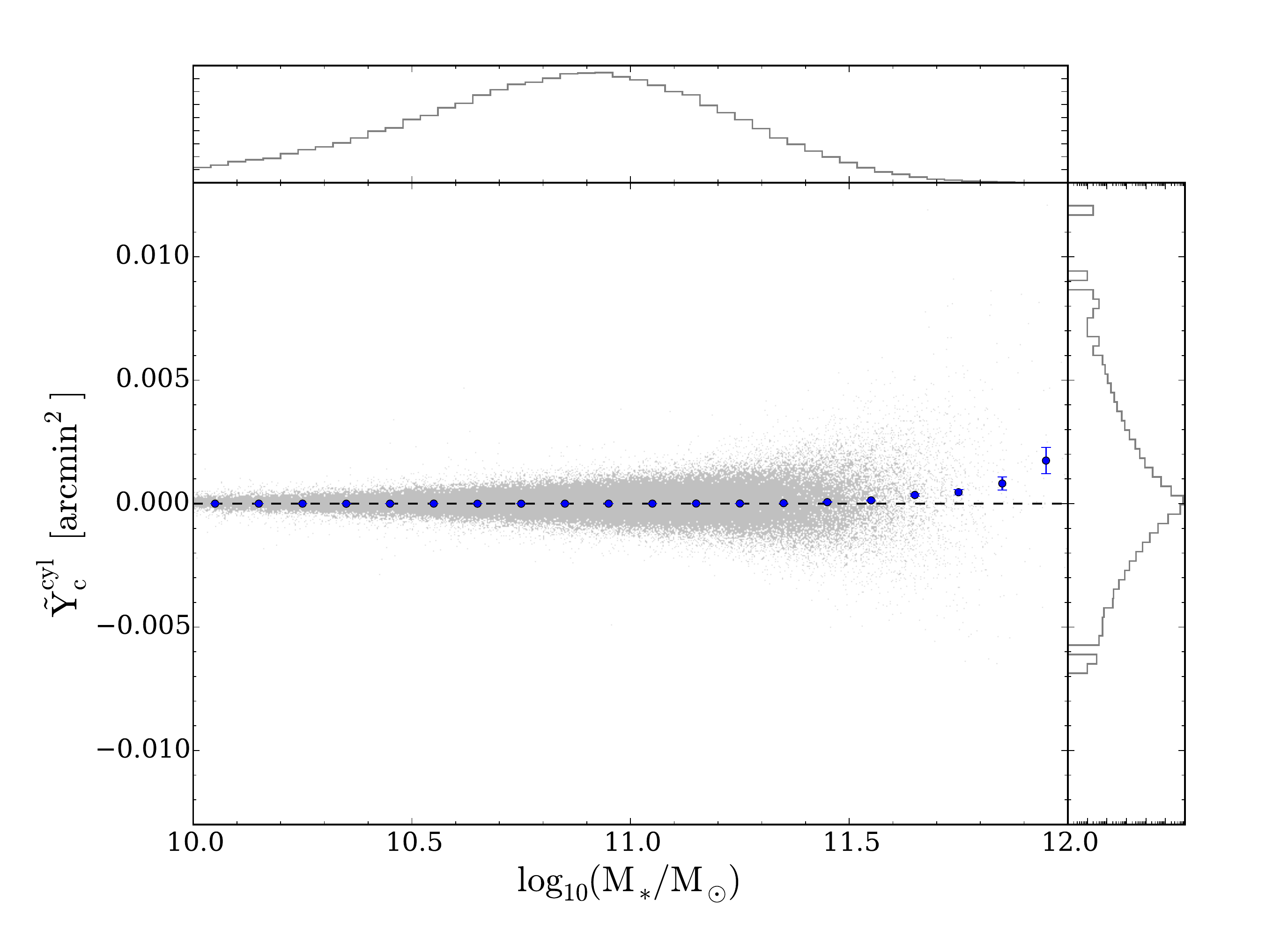}
\centering
\caption{The tSZ signal from individual LBGs (gray points) and stacked on stellar mass (blue circles). Note the slight asymmetry toward positive $\Yc$ values in the vertical histogram, which provides evidence that the data contain tSZ signal. Error bars represent $1\sigma$ uncertainties.}
\label{fig:Yc_all}
\end{figure}

\begin{figure*}[t!]
\centering
\includegraphics[width=15.15cm,trim=0.cm 0.2cm 0.cm 0.1cm,clip=true]{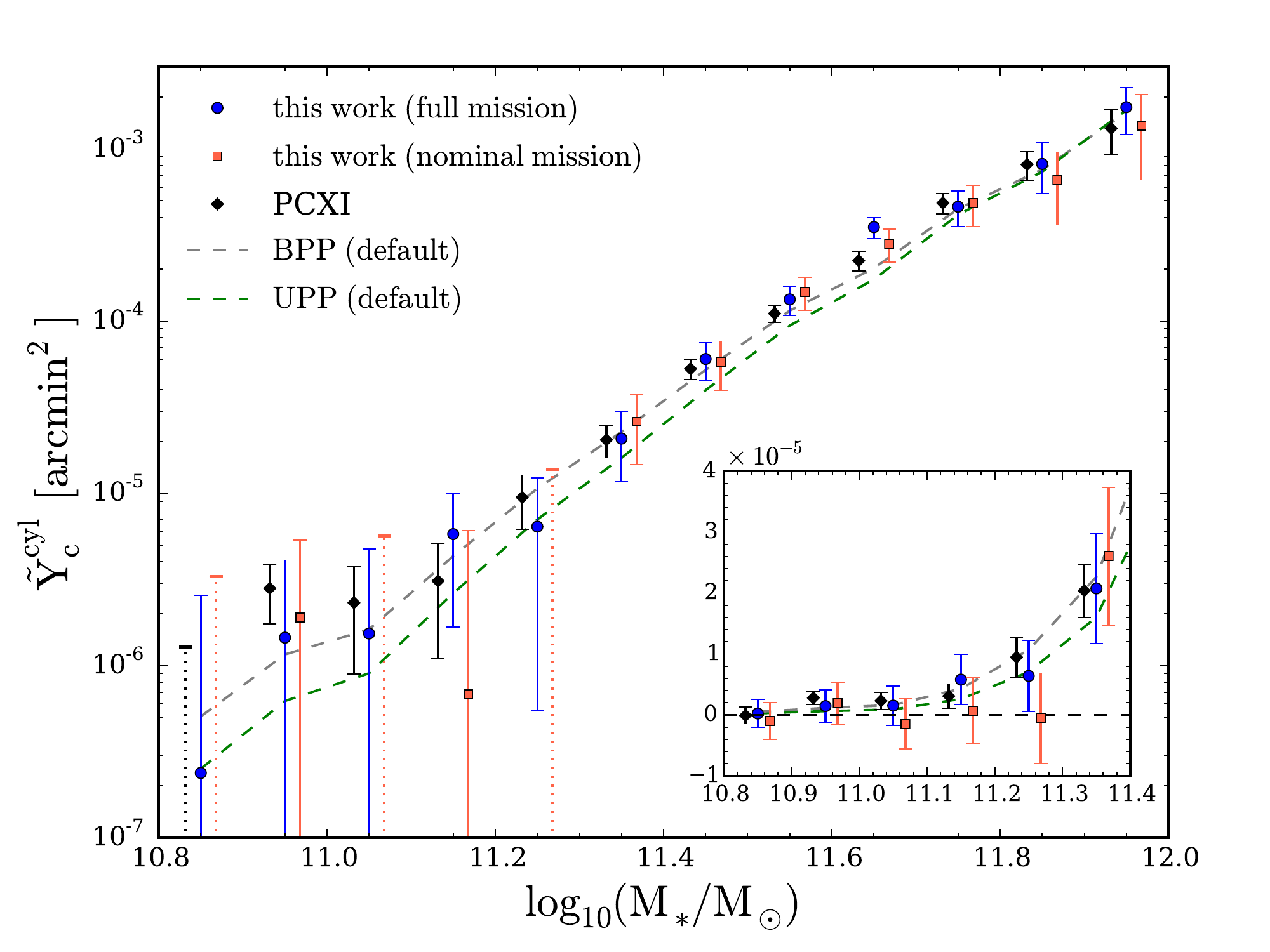}
\caption{The stacked tSZ signal from LBGs. We show the results of our fiducial analysis using the Planck nominal mission (orange squares) and full mission (blue circles) maps. The dashed lines indicate the predictions of the UPP and BPP, assuming our default model parameters. The theoretical calculations were performed by assuming the ``effective'' halo masses from Table~B.1 of PCXI and the observed median redshift for each stellar mass bin. For comparison with PCXI, we use the UPP to convert their $\tilde{Y}^\m{sph}_{500}$ into $\Yc$. Error bars represent $1\sigma$ uncertainties, and log-scale error bars with dotted-lines indicate bins with a negative stacked signal. We show $\chi^2$-test results in Table~\ref{deluxetable:fits}.} 
\label{fig:Y-Mstar}
\end{figure*}

Based on our fiducial, 4-channel analysis, Figure~\ref{fig:Yc_all} shows the tSZ signal from individual LBGs (gray points) and stacked on stellar mass (blue circles). The stacked signal is binned with 20 stellar mass bins in the range $10^{10}-10^{12}\, M_\odot$, and we estimate the binned averages and uncertainties with 30,000 bootstrap realizations per bin. This plot demonstrates the necessity for stacking in this analysis; the tSZ signal from our catalog of LBGs is a small statistical effect that is buried beneath an overwhelming amount of noise. Note the slight asymmetry toward positive $\Yc$ values in the vertical histogram, suggesting the data do indeed contain tSZ signal. In Figure~\ref{fig:Y-Mstar}, we compare of our fiducial analysis with the predictions of the default UPP and BPP models. These predictions are limited to the stellar mass bins for which PCXI provides a corresponding halo mass probability distribution function (Figure~B.1 of PCXI), which they generate from mock LBG catalogs based on the semi-analytic galaxy formation simulation of \citet{Guo2011}. We show the results of our analysis using the Planck nominal mission (orange squares) and full mission (blue circles) maps for comparison. Note the overall consistency of the results, with the full mission maps yielding higher signal-to-noise for all stellar masses. 

The theoretical predictions vary significantly with the assumed stellar-to-halo mass relation. For the calculations presented in this paper, we use the ``effective'' halo masses given in Table~B.1 of PCXI. These masses are ``measured'' from the PCXI mock LBG catalogs and account for aperture and miscentering effects, assuming the simulations accurately describe the spatial distribution of LBGs with respect to their host halos. If instead we simply integrate over the full halo mass probability distribution function for each stellar mass bin, we find the predictions are typically $1-2$ orders of magnitude higher than the effective mass predictions. We do not attempt to estimate the $Y-M_{500}$ relation because of the highly discrepant predictions from different $M_*-M_{500}$ methods. For the redshift dependence of the theoretical calculations, we use the observed median redshift for each stellar mass bin. 

 % **** Yc (primary results)
%****
% **** Yc (stacked)
%****

In order to compare our results with PCXI, we convert their $\tilde{Y}^\m{sph}_{500}$ to $\Yc$ with the UPP conversion factor given in \textsection\ref{sec:integratedtSZ}. In general, our results have lower signal-to-noise than those of PCXI. This is likely due to a combination of differences in our analyses. In particular, we smooth all maps to the resolution of the 100~GHz map, and this was not necessary in the analysis of PCXI, which allowed them to retain the full sensitivity of all channels. In addition, their signal extraction method utilized knowledge of the spatially inhomogeneous noise patterns in the maps to down-weight objects with particularly noisy signals. It is also worth mentioning that we sacrifice some signal-to-noise by applying a more conservative mask, which removes $\sim50\%$ of the most contaminated sky. 

We see clear evidence of the mean tSZ signal from LBGs with $\log_{10}(M_*/M_\odot) > 11.3$, and there is less significant evidence of the signal down to $\log_{10}(M_*/M_\odot) \sim 11.1-11.3$. For lower stellar masses, the signal is consistent with zero. As evidenced by the $\chi^2$-test results in Table~\ref{deluxetable:fits}, both the UPP and BPP predictions are in good agreement with the data. The large uncertainties in our measurements, particularly in the low-mass bins, make it impossible to rule out one of these models in favor of the other. Formally, the BPP provides a better fit to the data than the UPP, predicting slightly less signal from high-mass LBGs and more signal from low-mass LBGs --- the latter trend is favored by the data. 

What is the physical justification for these trends? For $r<R_{500}$, the UPP is an empirical profile, and in this radial range, the BPP agrees quite well with it. However, for $r>R_{500}$, the UPP is based on older simulations that over cool and remove too much gas from the ICM, which causes the pressure profile to be steeper than what is observed \citep{Planck_V}. In the simulations of \citet{Battaglia2010}, which are the basis of the BPP, feedback from AGN helps prevent overcooling and shuts off star formation across all halo masses, leading to higher ICM gas fractions and more pressure at larger radii; this is likely the reason for the trend seen at low stellar masses. At the high-mass end, the reason why the BPP predicts less signal than the UPP is not so clear. It may be related to the inherent hydrostatic mass bias in the UPP measurement (as noted in \textsection\ref{sec:tSZE}, we assume $(1-b)=0.8$ for all masses). 

One of the most surprising results from PCXI is the self-similarity in the $Y-M_{500}$ relation over more than two orders of magnitude in halo mass. While we do not attempt to estimate the $Y-M_{500}$ relation due to the aforementioned uncertainties associated with the stellar-to-halo mass relation, we nevertheless would like to test our results for evidence of deviation from self-similarity. To this end, we use the self-similar model parameters of the UPP and BPP (see \textsection\ref{sec:tSZE}) to calculate purely self-similar predictions of $\Yc$ (i.e., $\Yc(M_{500})\propto M_{500}^{5/3}$). We then vary the power-law index as $5/3 + \alpha$, where $-1<\alpha<1$, and for each model, we find the value of $\alpha$ that minimizes $\chi^2$ with respect to our fiducial analysis. Note these calculations are carried out in the $\Yc-M_*$ plane, where we use an underlying self-similar relation for the pressure profile's dependence on $M_{500}$. Similar to the theoretical calculations in Figure~\ref{fig:Y-Mstar}, we use the ``effective'' halo masses from Table~B.1 of PCXI for each stellar mass bin. In addition to the stellar-to-halo mass relation, these ``effective'' masses account for effects such as miscentering and satellite contamination. The results from both pressure profiles are shown in Table~\ref{deluxetable:alpha}. For comparison, the UPP and BPP with default parameters predict $\alpha=0.12$ and $\alpha\approx0.05$, respectively. Within the uncertainties of our measurements, $\alpha$ is consistent with zero in all cases. Thus, in the mass range where we detect tSZ signal, our results are consistent with an ICM pressure profile that scales self-similarly with mass. However, this result is very sensitive to scatter in the $M_*-M_{500}$ relation. If we use the mean halo masses given in Table~B.1 of PCXI rather than the ``effective'' halo masses, we find $\alpha{\sim}0.5$ for both the UPP and BPP. Thus, the astrophysical uncertainties associated with converting stellar masses to halo masses lead to an order-of-magnitude-level uncertainty in the conclusion about self-similarity. More precise data and a better understanding of the stellar-to-halo mass relation are needed to detect the relatively small deviations from self-similarity that are predicted by 
the UPP and BPP. 

\begin{deluxetable}{cccc}
\tablewidth{7.6cm}
\tablecolumns{4}
\tablehead{
\colhead{Pressure Profile}&
\colhead{$\chi^2$} &
\colhead{DOF} &
\colhead{PTE}
}
\startdata
UPP   &   18.5    & 12    &  0.10 \\
BPP  &   10.8  & 12    &  0.55 
\enddata
\tablecomments{The $\chi^2$, Degrees of Freedom (DOF), and Probability to Exceed (PTE) values from the fits of the UPP and BPP (default parameter) predictions to the data in Figure~\ref{fig:Y-Mstar}.
\vspace{-0.5cm}}
\label{deluxetable:fits}
\end{deluxetable}
\begin{deluxetable}{cc}
\tablewidth{7.6cm}
\tablecolumns{2}
\tablehead{
\colhead{Pressure Profile}&
\colhead{$\alpha$} 
}
\startdata
UPP   &    $-0.05_{-0.05}^{+0.06}$  \\
BPP   &    $-0.04_{-0.06}^{+0.06}$  
\enddata
\tablecomments{Best-fit values and uncertainties of the parameter $\alpha$, which is a measure of deviation from self-similarity. This parameter is derived from the purely self-similar predictions of the UPP and BPP, where we allow the power-law in mass to vary such that $\Yc(M_{500})\propto M_{500}^{5/3+\alpha}$. For reference, the UPP and BPP with default parameters predict $\alpha=0.12$ and $\alpha\approx0.05$, respectively.}
\label{deluxetable:alpha}
\end{deluxetable}

% **** Y500
\begin{figure}[h!]
\centering
\includegraphics[width=9.0cm, trim=0.25cm 0.cm 0cm 0cm,clip=true]{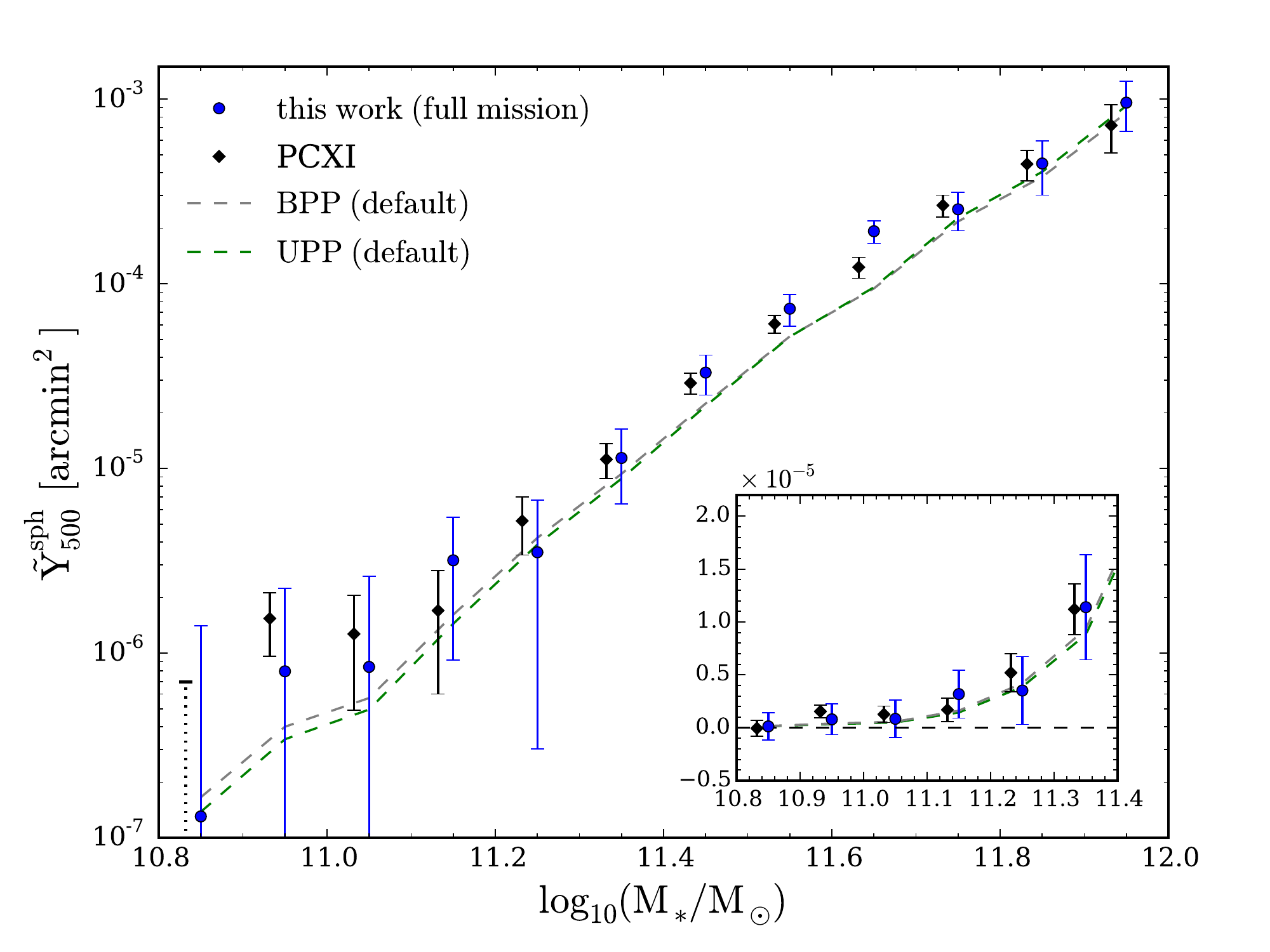}
\caption{Comparison of our fiducial analysis with the results of PCXI and the predictions of the UPP and BPP. In order to directly compare our results with PCXI, we use the UPP to convert our observed $Y^\m{cyl}_c$ into $Y^\m{sph}_{500}$. In practice, this means the data and UPP predictions have been scaled with respect to Figure~\ref{fig:Y-Mstar} by the same constant factor (see \textsection\ref{sec:integratedtSZ}). The BPP predictions in this figure were calculated by performing the spherical integration in Equation ~(\ref{eqn:Y500}) for each stellar mass bin. Note we have neglected the scatter in the $Y^\m{cyl}_c - Y^\m{sph}_{500}$ plane (see Figure~\ref{fig:scatter}), so the uncertainties in this figure are underestimated. Error bars represent $1\sigma$ uncertainties, and log-scale error bars with dotted-lines indicate bins with a negative stacked signal. 
The larger discrepancy between the data and the BPP in this figure as compared to Figure~\ref{fig:Y-Mstar} highlights the limitations of $Y^\m{sph}_{500}$ as a profile-dependent quantity (see Figure~\ref{fig:scatter} and \citet{LeBrun2014}). \label{fig:Y500}
\vspace{0.3cm}}
\end{figure}
% **** 

As mentioned in \textsection\ref{sec:integratedtSZ}, observations directly probe $Y^\m{cyl}_c$, but $Y^\m{sph}_{500}$ is more strongly correlated with cluster mass, making it a theoretically interesting quantity. Converting $Y^\m{cyl}_c$ into $Y^\m{sph}_{500}$, however, requires knowledge of the electron pressure profile, and this makes $Y^\m{sph}_{500}$  non-optimal for constraining the pressure profile itself.  For direct comparison with PCXI, we convert $Y^\m{cyl}_c$ into $Y^\m{sph}_{500}$ by assuming the UPP is valid over the mass and redshift ranges of our sample. Figure~\ref{fig:Y500} shows the results along with the predictions of the UPP and BPP. In practice, the data and UPP predictions have been scaled with respect to Figure~\ref{fig:Y-Mstar} by the same constant factor (see \textsection\ref{sec:integratedtSZ}). The BPP conversion, however, depends on both mass and redshift, and its predictions were calculated by integrating over a sphere of radius $R_{500}$ (Eqn.~(\ref{eqn:Y500})) for each stellar mass bin, where we have assumed the same halo masses and redshifts as in Figure~\ref{fig:Y-Mstar}. The larger discrepancy between the data and the BPP in this figure as compared to Figure~\ref{fig:Y-Mstar} is an indication that converting  $Y^\m{cyl}_c$ into $Y^\m{sph}_{500}$ is non-optimal for constraining the ICM pressure profile --- $Y^\m{sph}_{500}$ is not an observable quantity  (see Figure~\ref{fig:scatter} and \citet{LeBrun2014}). Within the uncertainties, our results are in very good agreement with those of PCXI. Note the errors in this figure are underestimated, as we have neglected the scatter in the $Y^\m{cyl}_c-Y^\m{sph}_{500}$ plane (see Figure~\ref{fig:scatter}).

\subsection{Sensitivity to Dust Model}\label{sec:dust_test}
\begin{figure}[t!]
\centering
\includegraphics[width=9.0cm, trim=0.25cm 0.2cm 0cm 0cm,clip=true]{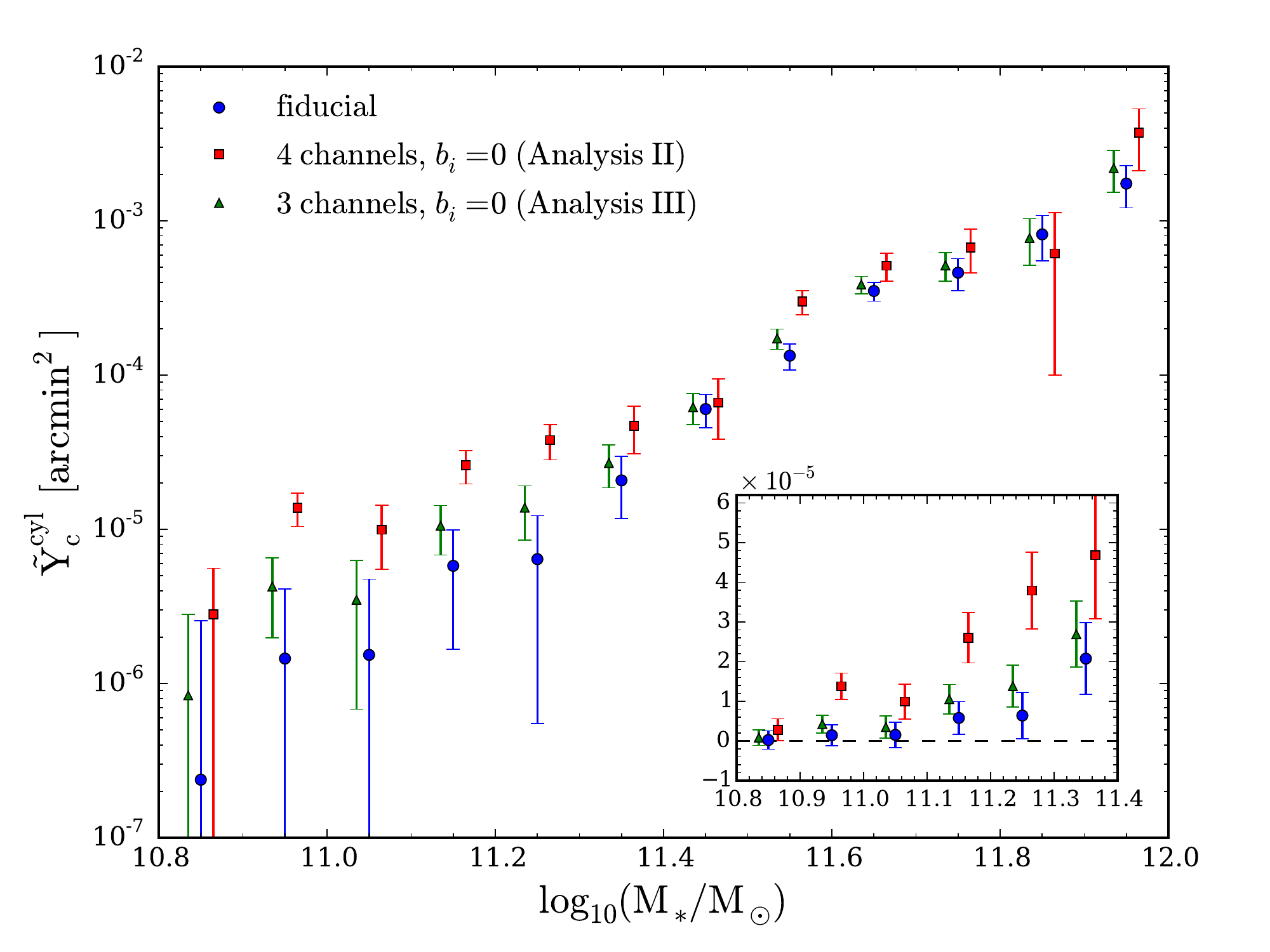}
\caption{This figure compares our fiducial analysis with Analyses II and III (see \textsection\ref{sec:extract}), both of which do not model dust emission. The excess signal seen in Analysis II is an indication that dust contamination is an issue for this analysis, especially for the low-mass LBGs. Analysis III is expected to contain very little signal from dust. However, it is biased high with respect to our fiducial analysis, and this suggests it does in fact contain non-negligible dust contamination. Error bars represent $1\sigma$ uncertainties. \label{fig:compare}}
\end{figure}

In our fiducial analysis,  $b_i$  (Eqn. (\ref{eqn:model})) accounts for dust emission associated with the target LBGs and their host halos. We test our assumptions about the frequency and redshift dependence of this emission in Figure~\ref{fig:compare}, which compares the stacked tSZ signal from our fiducial analysis with our four and three channel analyses with $b_i=0$ (Analysis II and Analysis III, respectively). In contrast to PCXI, we see clear evidence for dust emission at all stellar masses. In particular, Analysis II shows significant excess signal with respect to the other analyses. By excluding the highest frequency channel, Analysis~III is much less sensitive to residual dust emission and provides a useful test for our assumed dust model. However, it is biased high with respect to our fiducial analysis, and this suggests it does in fact contain non-negligible dust contamination, likely in the 217 GHz channel. Our fiducial analysis appears to successfully suppress the excess signal seen in Analyses II and III while utilizing more frequency information to probe both the tSZ and dust signals.

% **** Yc model compare
%****

\subsection{The stacked dust signal}\label{sec:dustsig}

\begin{figure}[h!]
\begin{minipage}[h!]{0.5\linewidth} 
\centering
\includegraphics[width=8.5cm,trim=0.28cm 0.3cm 0.cm 0.5cm,clip=true]{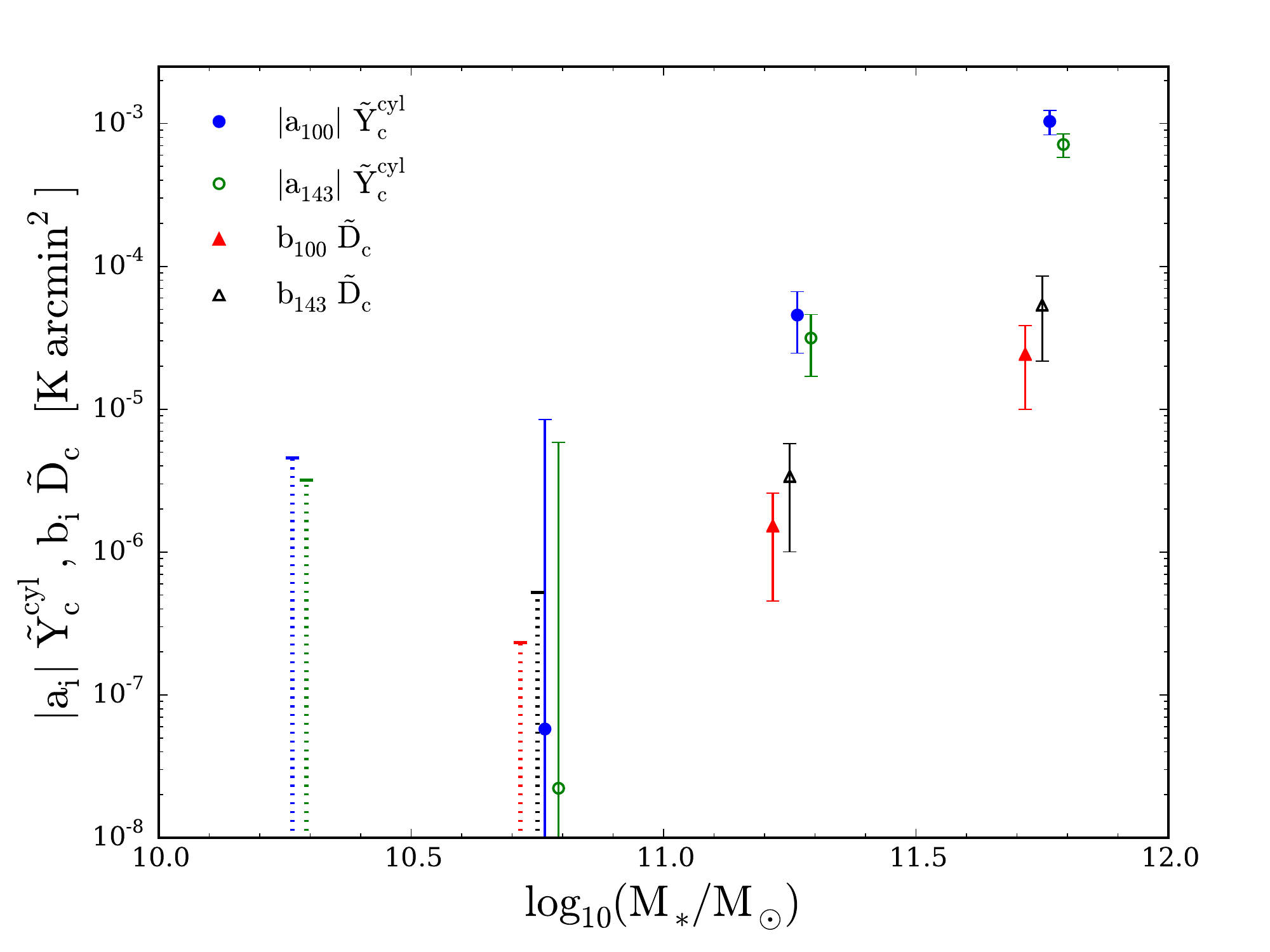}
\end{minipage}
\begin{minipage}[h!]{0.5\linewidth}
\centering
\includegraphics[width=8.5cm,trim=0.28cm 0.3cm 0.cm 0.3cm,clip=true]{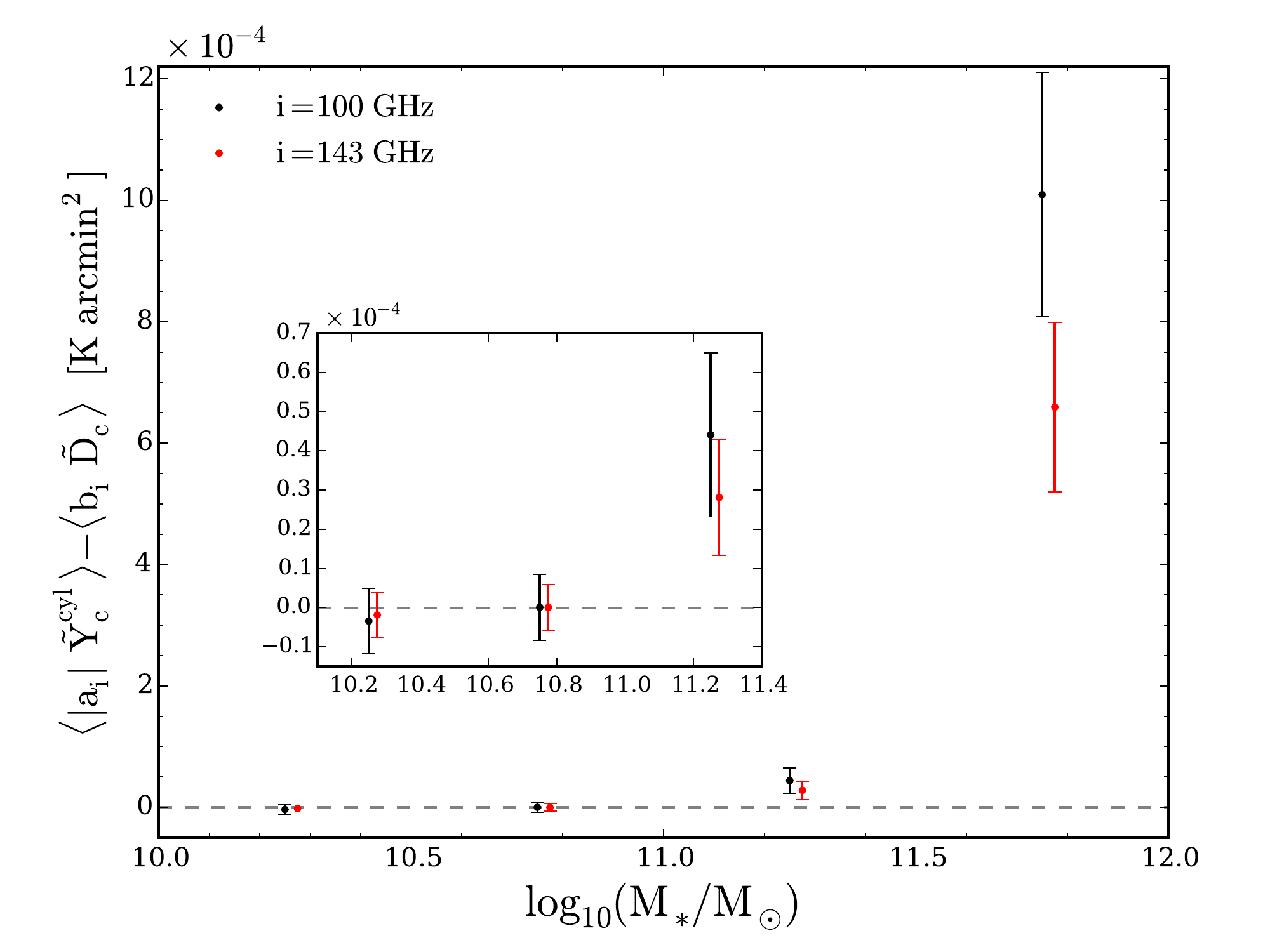}
\end{minipage}
\caption{The top panel compares the stacked dust ($b_i \tilde{D}_c$) and tSZ ($|a_i|\Yc$) signals in the $i = 100$ and $143$~GHz channels (see Equation~(\ref{eqn:model})). The bottom panel is a measure of the relative importance of dust contamination with respect to the tSZ signal in these channels, showing that dust is a significant contaminant for the low-mass LBGs (i.e., $\langle b_i \tilde{D}_c\rangle \gtrsim \langle |a_i| \Yc\rangle$) and contributes non-negligible contamination to the observed signal of high-mass LBGs. In both panels, error bars represent $2\sigma$ uncertainties, and log-scale error bars with dotted-lines indicate bins with a negative stacked signal. For the stacked dust signal in the lowest-mass bin, the $2\sigma$ error bar is negative and hence does not appear on the log-scale plot.\label{fig:Dc}}
\end{figure}

The dust parameter, $D_c$, probes the integrated dust emission from the target LBGs and their parent halos. As Figure~\ref{fig:compare} suggests, dust can be a significant contaminant for tSZ measurements, and it is therefore important to simultaneously constrain both emission from dust and the tSZ signal. The top panel of Figure~\ref{fig:Dc} compares the stacked dust ($b_i \tilde{D}_c$) and tSZ ($|a_i|\Yc$) signals in the $i = 100$ and $143$~GHz channels (see Equation~(\ref{eqn:model})). The data are plotted in 4 stellar mass bins, where the binned averages and uncertainties are estimated with 10,000 bootstrap realizations per bin. We see the total amount of dust emission from LBGs increases with stellar mass, which is sensible over the mass range of interest. The bottom panel of this figure is a measure of the relative importance of dust emission with respect to the tSZ signal, expressed with the quantity $\langle\ |a_i| \Yc\rangle - \langle b_i \tilde{D}_c\rangle$, where $\langle...\rangle$ denotes the average over a given stellar mass bin. This plot makes it clear that dust is a significant contaminant for the low-mass LBGs (i.e., $\langle b_i \tilde{D}_c\rangle \gtrsim \langle |a_i| \Yc\rangle$), but dust also contributes non-negligible contamination to the observed signal of high-mass LBGs. 

% **** D500
% **** 

\subsection{Sensitivity to miscentering}\label{sec:miscenter}

Miscentering between the LBGs and their host halos broadens the mean tSZ profile by pushing flux into the tail of the distribution, and for significant offsets, the integrated signal will be suppressed due to the finite aperture size of the observation. PCXI investigate the impact of miscentering with their mock LBG catalog, and in Table~C.1, they tabulate mean and RMS offsets of LBGs from the gravitational potential minima of their parent halos. It is important to emphasize that the predictions of the UPP and BPP presented in this paper use the ``effective'' halo masses from Table B.1 of PCXI, which account for aperture and miscentering effects. Here, we present a simple empirical test for the effects of miscentering. 

The impact of miscentering on our analysis should be more pronounced at low redshifts since a given physical offset corresponds to a larger angular separation on the sky at low-z. If our results are sensitive to this effect, we expect, for fixed stellar mass, the tSZ signal of low-redshift LBGs will be biased low with respect to their high-redshift counterparts. We, therefore, test our sensitivity to miscentering with the quantity $\Delta \Yc \equiv\, \langle\Yc(z>z_\m{med})\rangle - \langle\Yc(z<z_\m{med})\rangle$, where $z_\m{med}$ is the median redshift of LBGs in a given stellar mass bin. Note that $\Yc$ scales out the self-similar evolution, so any differences arising from redshift evolution should be quite small. We find that $\Delta \Yc$ is consistent with zero, using 20 stellar mass bins, as in our fiducial analysis, and using  3 bins, which increases the signal-to-noise in the high-mass bins. It may be that the redshift ranges of each bin are not large enough to see this effect. It is also possible that the signal is intrinsically lower for a reason other than miscentering, which would misleadingly yield results that are consistent with the predictions of the UPP and BPP. 

\section{Conclusions} \label{sec:conclusion}

In this work, we have presented a measurement of the stacked tSZ signal from locally brightest galaxies selected from SDSS/DR7. This study was motivated by the potentially surprising findings of PCXI, which suggest non-gravitational processes do not have a strong impact on the thermodynamic state of hot gas in low-mass halos. While our analysis closely follows that of PCXI, it differs in several important ways. Most significantly, we explicitly treat dust emission from each LBG in our fiducial analysis, and we measure a stacking induced bias and subtract it from our results (for details see \textsection\ref{sec:bias}). This stacking bias becomes significant for the lowest-mass halos probed by this study. 

The primary results of this paper can be summarized as follows:
\begin{itemize}
  \item We report a significant measurement of the stacked tSZ signal from LBGs with $\log_{10}(M_*/M_\odot)>11.3$, with some evidence of the signal down to $\log_{10}(M_*/M_\odot)\sim11.1-11.3$  (Figure~\ref{fig:Y-Mstar}). This result is consistent with the findings of PCXI.  
    \item The stacked signal from dust emission is comparable to or larger than the stacked tSZ signal from LBGs with $M_* \lesssim 10^{11.3}\,M_{\odot}$ (Figure~\ref{fig:Dc}). Above this stellar mass, we find dust emission contributes non-negligible contamination to the observed signal, which is contrary to the findings of PCXI. 
      \item The BPP provides a formally better fit to our results than the UPP (Table~\ref{deluxetable:fits}). However, it is important to point out that uncertainties in both the $M_*-M_{500}$ relation and our measurements make it impossible to rule out one of these models in favor of the other.  
   \item Within the uncertainties of our measurements, our results are consistent with a self-similar ICM pressure profile down to the lowest mass scales for which we detect the tSZ signal (Table~\ref{deluxetable:alpha}). 
\end{itemize}
We emphasize that more precise measurements, larger group/cluster  samples, and a better understanding of the stellar-to-halo mass relation or direct lensing mass estimates for these objects are needed to make definitive statements about the self-similarity of the gas pressure profiles of low-mass halos. Fortunately, upcoming data from future small-scale CMB experiments such as ACTPol \citep{ACTPol2010}, Advanced ACTPol \citep{Calabrese2014}, SPTpol \citep{SPTpol2012}, and SPT-3G \citep{Benson2014} should have the necessary sensitivity to probe deviations from self-similarity, providing fundamental constraints on the importance of feedback in the formation and evolution of galaxies.

\begin{acknowledgments}
JPG is supported by the National Science Foundation Graduate Research Fellowship under Grant No. DGE 1148900. JCH and DNS acknowledge support from NASA Theory Grant NNX12AG72G and NSF AST-1311756. JCH also gratefully acknowledges funding from the Simons Foundation through the Simons Society of Fellows. We greatly appreciate very useful comments from Simon White and Jean-Baptiste Melin. JPG is grateful for useful conversations with Michael Strauss. JCH would like to thank Michael Anderson and Amandine le Brun for useful conversations.
\end{acknowledgments}

\bibliographystyle{apj}
\bibliography{mybib}

\end{document}